\documentclass[10pt,letterpaper]{article}
\usepackage{opex3}
\usepackage{amssymb,amsmath} 
\usepackage{cite}

\begin{document}
\title{Interplay between localization and absorption in disordered waveguides}
\author{Alexey~G.~Yamilov$^{*}$ and  Ben~Payne}
\address{Department of Physics, Missouri University of Science \& Technology, Rolla, MO 65409}
\email{$^{*}$yamilov@mst.edu} 

\begin{abstract} 
This work presents results of ab-initio simulations of continuous wave transport in disordered absorbing  waveguides. Wave interference effects cause deviations from diffusive picture of wave transport and make the diffusion coefficient position- and absorption-dependent. As a consequence, the true limit of a zero diffusion coefficient is never reached in an absorbing random medium of infinite size, instead, the diffusion coefficient saturates at some finite constant value. Transition to this absorption-limited diffusion exhibits a universality which can be captured within the framework of the self-consistent theory (SCT) of localization. The results of this work (i) justify use of SCT in analyses of experiments in localized regime, provided that absorption is not weak; (ii) open the possibility of diffusive description of wave transport in the saturation regime even when localization effects are strong.
\end{abstract}




\section{Introduction\label{sec:intro}}

The diffusive description of electromagnetic wave propagation in inhomogeneous media\cite{1953_Morse,1999_van_Rossum} is overwhelmingly successful with wide ranging applications from astrophysics\cite{1960_Chandra} to biomedical optics\cite{2007_Wang_Biomedical_Optics}. Wave interference effects are usually negligible because the phases of the multiply scattered partial waves are random. However, for a wave that returns to its source location there is always the time-reversed path which necessarily yields the same phase. When the return probability is sufficiently high, the constructive interference between the wave propagated via the two paths can suppress long-range transport so that the diffusion coefficient in an infinitely large (passive) system turns to zero\cite{1980_Vollhardt_Wolfle,1993_Kroha_self_consistent}. This is Anderson localization -- the phenomenon first conceived in context of electronic (de Broglie) wave transport in condensed matter physics\cite{1958_Anderson}. 

The question of how the diffusion coefficient evolves with an increase of system size from the unrenomalized value of $D_0$ toward the limit of $D=0$ has been addressed in Ref.~\cite{2000_van_Tiggelen,2008_Cherroret} by extending the original self-consistent theory of Vollhardt and W\"olfle\cite{1980_Vollhardt_Wolfle,1993_Kroha_self_consistent} to systems of finite size. The key prediction of the modified self-consistent theory (SCT) is that the diffusion coefficient is no longer a constant but varies spatially. This conclusion is also reached independently in the supersymmetric field theory\cite{2008_Tian}. The position dependence of the diffusion coefficient arises because the return probability responsible for the renormalization of diffusion is position-dependent in a system of finite size. As the wave explores the larger and larger neighborhood of a source point, the return probability becomes sensitive to the proximity of a boundary where the wave has a chance to escape from the system. In fact, Ref.~\cite{2013_Tian_review} points out that waves propagation in open media can be thought as a highly unconventional macroscopic diffusive phenomenon.

Although SCT has been successful in interpreting the results of several experiments\cite{2006_Maret_PRL,2008_van_Tiggelen_Nature,2009_Genack_PRB,2013_Maret_3D_localization}, the key prediction of position-dependent diffusion has not been verified in an experiment because it requires the access to the inside of the random medium. Position-dependent diffusion has been also predicted by the super-symmetry approach\cite{2008_Tian}, suggesting that this is more than just a convenient mathematical abstraction. Indeed, quantitative agreement between SCT and numerical simulation has been found in disordered waveguides at the onset of localization\cite{2010_Payne_PRL,2013_Yamilov_Closed_Channels}. Numerical simulations in Ref.~\cite{2010_Tian_PRL}, show that deeper in the localization regime, the prediction by SCT for the position-dependent diffusion coefficient is quantitatively incorrect. Instead, by using the supersymmetric field theory it was found that the diffusion coefficient exhibits novel scaling\cite{2010_Tian_PRL}. This is because SCT underestimates the energy density inside the bulk of medium that is strongly affected by the presence of necklace states\cite{1994_Pendry} formed via resonant tunneling.

Material absorption and other sources of dissipation are usually a nuisance in experiments with classical waves because it leads to the same exponential scaling of e.g. optical conductance $g\propto\exp(-L/\xi_{a0})$ as in a localized system $g\propto\exp(-L/\xi)$\cite{2000_chabanov_nature}. Here $\xi_{a0}=\sqrt{D_0\tau_a}$ is the diffusive absorption length, $\xi$ is the localization length, $\tau_a$ is the ballistic absorption time, and $g$ is the total transmission through random medium under diffuse illumination\cite{2000_chabanov_nature}. When the absorption rate $1/\tau_a$ exceeds the radiative decay rate through the boundaries of the random medium, the resonant tunneling through the modes of the random  medium\cite{2011_Genack_Nature} is suppressed. Because the radiative decay rate diminishes exponentially with an increase of the system size $L$, the amount of absorption needed to suppress the resonant tunneling becomes small. 

In this work, we show that the applicability of SCT in the localized regime is restored when sufficiently strong ($\xi_{a0}\lesssim L$) absorption is added. We also study the size scaling of the position-dependent diffusion coefficient and show that in presence of absorption it saturates at a non-zero value. This suggests that that the wave transport in absorbing random media can be viewed as diffusion, albeit with a renormalized diffusion coefficient, even when localization corrections are present and strong.

\section{Model description\label{sec:model}}

\subsection{Motivation for the choice of the model}

The goal of this work is to study the effect of an interplay between absorption and localization on the position-dependent diffusion coefficient. To achieve this goal we selected the model geometry based on the following considerations. Because the diffusion coefficient is modified due to localization corrections, the system must be in the localization regime in the limit of infinitely large system size. Scaling theory of localization predicts\cite{1979_Anderson} that this condition is met in systems with physical dimension less than or equal to two. In three dimensional random media the disorder strength has to be sufficiently high to reach localization\cite{2013_Maret_3D_localization}. From a computational standpoint, it is desirable to consider a system with the least dimension. Although a one-dimensional (1D) random medium, e.g. a random stack of dielectric slabs\cite{2001_Deych_sps_abs} or a single mode random waveguide\cite{2010_Sapienza_Lodahl_1D_Localization}, exhibits the phenomenon of localization it is not suitable for our study. This is because wave transport in 1D shows either ballistic or localized behavior; diffusion is not applicable in any parameter range. Instead, we perform numerical simulations for 2D disordered multimode waveguides. Because the volume of the system increases as the first power of the waveguide length $L$ (for fixed width $W$), this geometry is often referred to as a quasi-1D\cite{1997_Beenakker,1998_Brouwer}. However, unlike 1D, random waveguides exhibit both a diffusive behavior for $\ell<L<\xi$ and localization for $L>\xi$. $\ell$ denotes the transport mean free path. Furthermore, as we argue in Sec.~\ref{sec:conclusions}, the 2D random waveguides can be fabricated experimentally\cite{2013_Yamilov_Dofz_experiment} to test the predictions of this work.

\subsection{Position-dependent diffusion coefficient}

To investigate the applicability of a diffusive description of wave transport, we perform the ab-initio numerical simulations where all wave interference effects are accounted for without any approximations. We consider a scalar, monochromatic wave $E(\mathbf{r})e^{-i \omega t}$ propagating in a 2D volume-disordered waveguide of width $W$ and length $L \gg W$ that supports $N$ propagating modes. The wave field $E(\mathbf{r})$ obeys the 2D Helmholtz equation:
\begin{equation}
\left\{\nabla^2 + k^2\left[1 + \delta\epsilon(\mathbf{r}) \right]\right\} E(\mathbf{r}) = 0.
\label{eq:helmholtz}
\end{equation}
Here $k=\omega/c$ is the wavenumber, $\delta\epsilon(\mathbf{r})=(1+i\alpha)\delta\epsilon_r(\mathbf{r})$, where $\delta\epsilon_r(\mathbf{r})$ is the randomly fluctuating part of the dielectric constant, and $\alpha>0$ is the strength of absorption. The system is excited from the left by illuminating the waveguide with $N$ unit fluxes and the wave field $E(\mathbf{r})$ throughout the volume of the random medium is computed with a transfer matrix method for a given realization of disorder\cite{2013_Yamilov_Closed_Channels}. From $E(\mathbf{r})$ we calculate the cross section-averaged energy density ${\cal W}(z)$ and the longitudinal component of flux $J_z(z)$. Here $z$ is the axis along the length of the waveguide. Local values of the energy density and flux formally define via Fick's law the diffusion coefficient $D(z)$ which, in general, may be position-dependent:
\begin{equation}
D(z) = -\langle J_z(z)\rangle / \left[ d\langle{\cal W}(z)\rangle/d z\right].
\label{eq:Dofz_definition}
\end{equation}
The averages $\langle \ldots \rangle$ are taken over a statistical ensemble of $10^6$ disorder realizations simulated on a supercomputer. 

\subsection{$D_0$ calculation}

In order to compare our numerical results for $D(z)$ with SCT without fitting parameters, we need to obtain the value of the diffusion coefficient unrenormalized by the wave interference effects $D_0=v\ell/2$. Here $v$ is the diffusive speed and $\ell$ is the transport mean free path.

To obtain the value of the transport mean free path in our model we perform a set of simulations for different waveguide lengths, exploring both the regime of classical diffusion $L<\xi$ and that of Anderson localization $L>\xi$, where $\xi=(\pi/2)N\ell$. The dependencies of the ensemble-averaged conductance $\langle g \rangle$ and its variance $\mathrm{var}(g)$ on the parameter $g_0=(\pi/2)N\ell/(L+2z_0)$ are fitted\cite{2013_Yamilov_Closed_Channels} by the analytic expressions obtained by Mirlin in Ref.~\cite{2000_Mirlin} using the supersymmetry approach. Here, $g_0$ the diffusive expression for the average conductance which has the property $g_0\simeq\langle g\rangle$ for $L\ll\xi$. The expression for $g_0$ has two parameters $\ell$ and the so-called extrapolation length $z_0$ \cite{1999_van_Rossum}that are used to fit numerical data. The presence of scatterers inside the waveguide causes the effective refractive index in the disordered region to be different from that in the empty regions. The refractive index mismatch between can cause surface reflections for the waves trying to escape from the system\cite{1999_van_Rossum}. In the absence of such reflections, $z_0$ is related to the transport mean free path as $z_0=(\pi/4)\ell$. Assuming this relationship between $z_0$ and $\ell$ we obtain a good fit. However, by adjusting $\ell$ on the order of $2\%$ and treating $z_0$ as an independent fitting parameter, we achieve even better agreement between the position dependent diffusion coefficient computed from the ab-initio numerical simulations and from self-consistent theory as reported in Sec.~\ref{sec:analysis} below. This suggests the existence of some surface reflection in the disorder model\cite{2013_Yamilov_Closed_Channels} we use in numerical simulations.

To find the diffusive speed $v$ we use the definition of diffusive flux resolved with respect to the direction of propagation\cite{1953_Morse} 
\begin{equation}
\langle J^{(\pm)}_z(z)\rangle = (v/\pi)\langle{\cal W}(z)\rangle \mp (D(z)/2)d\langle {\cal W}(z)\rangle/dz,
\end{equation}
where $J^{(\pm)}_z(z)$ represent the forward (plus) and backward (minus) propagating components of the flux. Combining the two components we find
\begin{equation}
v=\frac{\pi}{2}\displaystyle{\frac{\langle J^{(+)}_z(z)\rangle + \langle J^{(-)}_z(z)\rangle}{\langle{\cal W}(z)\rangle}}.
\label{eq:v}
\end{equation}

\subsection{$\tau_a$ calculation}

The characteristic absorption time $\tau_a$ can be determined numerically using the condition of flux continuity $d\left\langle J_z(z)\right\rangle/dz = \left\langle{\cal W}(z)\right\rangle/\tau_a$.  The desired diffusive absorption length $\xi_{a0}=\sqrt{D_0\tau_a}$ can be obtained by the proper choice of $\alpha$ in Eq.~(\ref{eq:helmholtz}).

\section{Self-consistent theory\label{sec:SCT}}

Self-consistent theory is developed starting with the Green's function $G(\mathbf{r}, \mathbf{r}')$ of Eq.~(\ref{eq:helmholtz}) with $\delta\epsilon(\mathbf{r})=\delta\epsilon_r(\mathbf{r})+i\alpha$. The disorder-averaged function ${\hat C}(\mathbf{r}, \mathbf{r}') = (4\pi W D_0/cL)\langle \left| G(\mathbf{r}, \mathbf{r}') \right|^2 \rangle$ obeys self-consistent equations in a dimensionless form\cite{2008_Cherroret,2010_Payne_PRL}: 
\begin{eqnarray}
&&\left[\left(\frac{L}{\xi_{a0}}\right)^2 - \frac{\partial}{\partial \zeta} d(\zeta)
 \frac{\partial}{\partial \zeta} \right] {\hat C}(\zeta,\zeta')
= \delta(\zeta-\zeta'),
\label{eq:sceq1}
\\
&&\frac{1}{d(\zeta)} =  1+\frac{2L}{\xi}
{\hat C}(\zeta,\zeta),
\label{eq:sceq2}
\end{eqnarray}
where $d(\zeta) = D(\zeta)/D_0$ and all position-dependent quantities are functions of the longitudinal coordinate $\zeta = z/L$ in the considered waveguide geometry. The quantity ${\hat C}(\zeta,\zeta)$ which renormalizes the diffusion coefficient is proportional to the return probability at $\zeta$. Hence, Eq.~(\ref{eq:sceq2}) formally accounts for wave interference of the time-reversed paths. Eqs.~(\ref{eq:sceq1},\ref{eq:sceq2}) are solved with the following  boundary conditions:
\begin{eqnarray}
{\hat C}(\zeta,\zeta^{\prime}) \mp
\frac{z_0}{L} d(\zeta) \frac{\partial}{\partial \zeta}
{\hat C}(\zeta,\zeta^{\prime}) = 0
\label{eq:bc}
\end{eqnarray}
at $\zeta = 0$ and $\zeta = 1$. We obtain the numerical solution of Eqs.~(\ref{eq:sceq1}-\ref{eq:bc}) via an iterative procedure with an initial guess $d(\zeta)\equiv 1$. The convergence is usually achieved within five iteration steps.

\section{Analysis of numerical results\label{sec:analysis}}

\subsection{Interplay between localization and absorption}

\begin{figure}
\vskip -1cm
\centering{\includegraphics[width=3in,angle=-0]{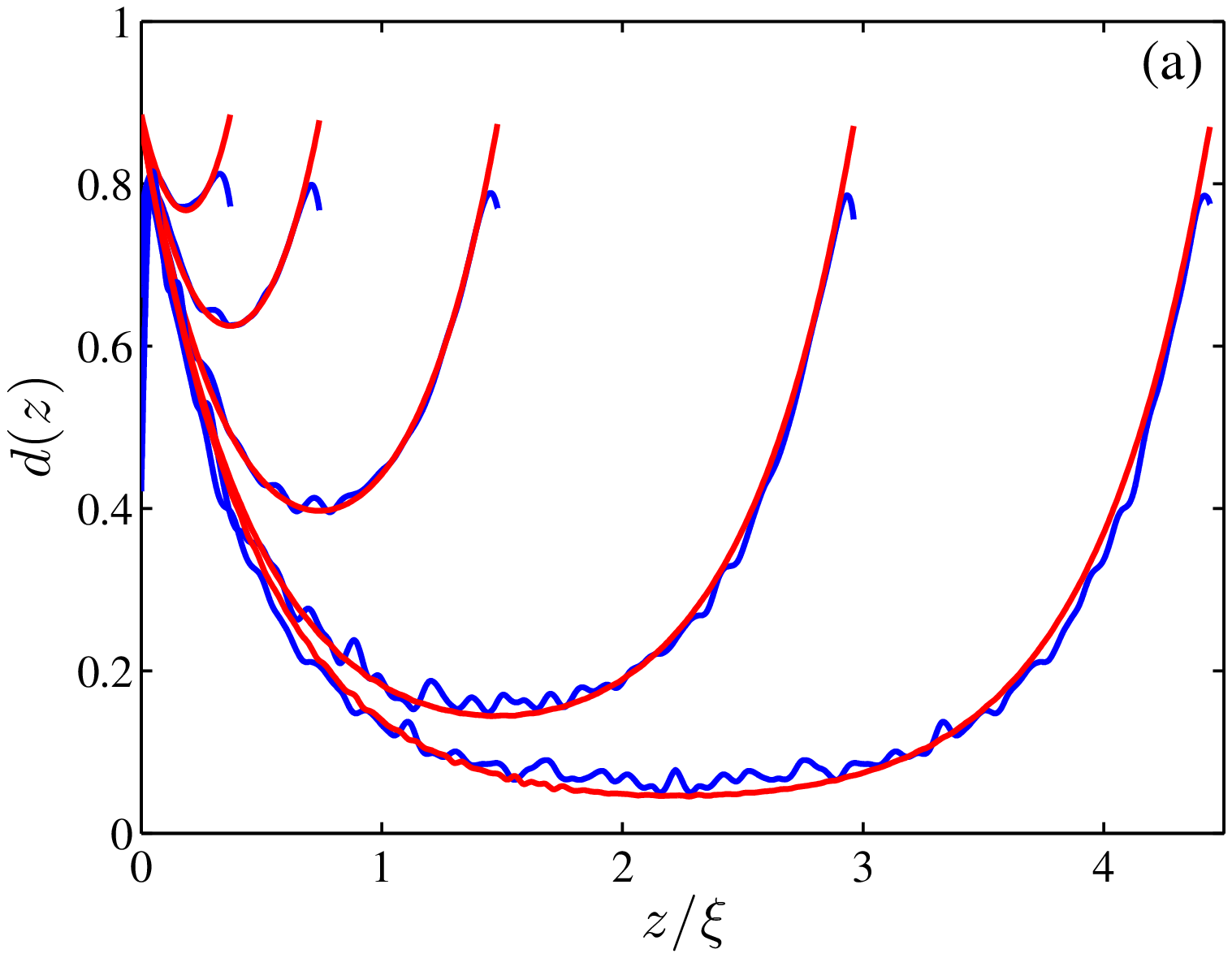}}
\centering{\includegraphics[width=3in,angle=-0]{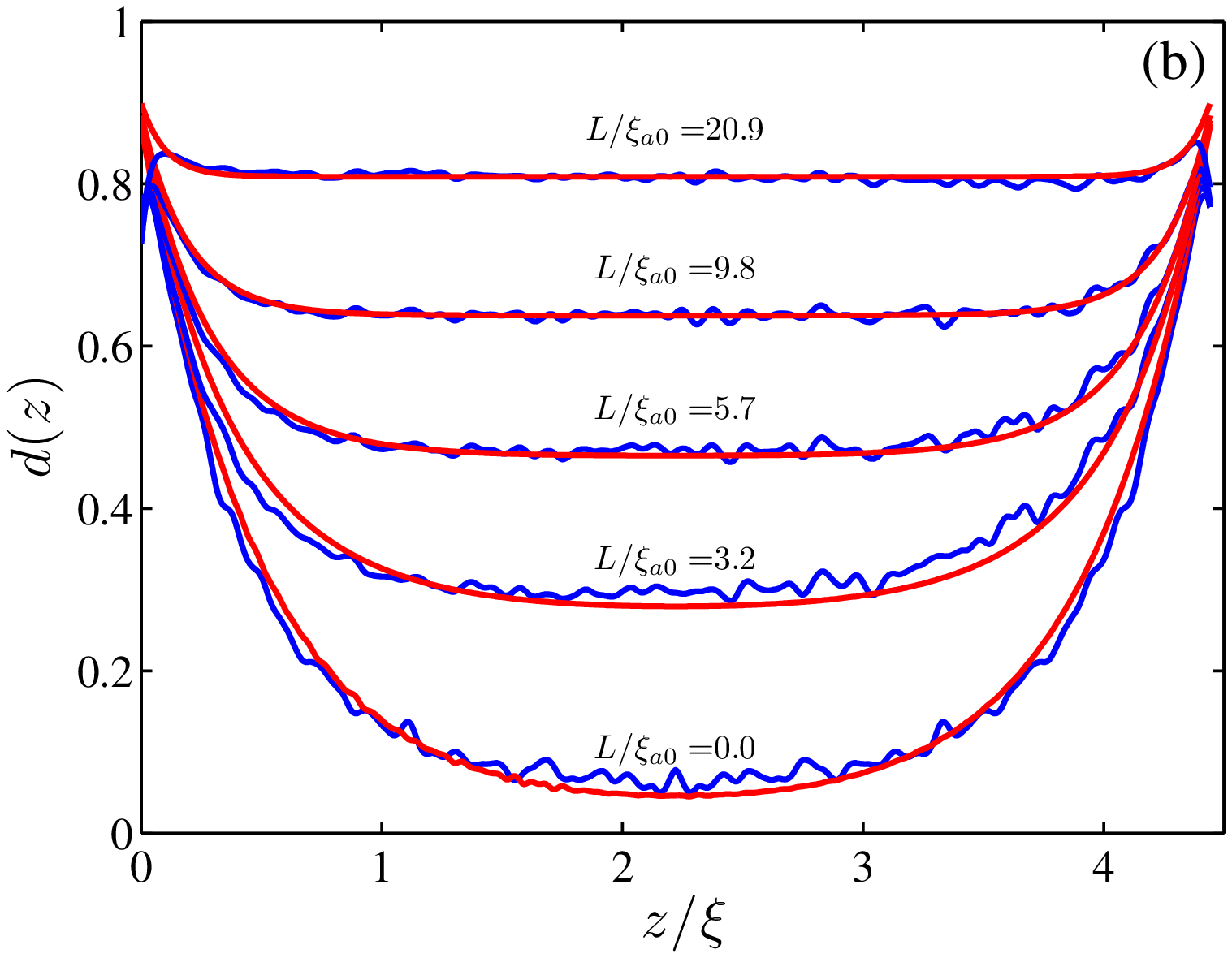}}
\centering{\includegraphics[width=3in,angle=-0]{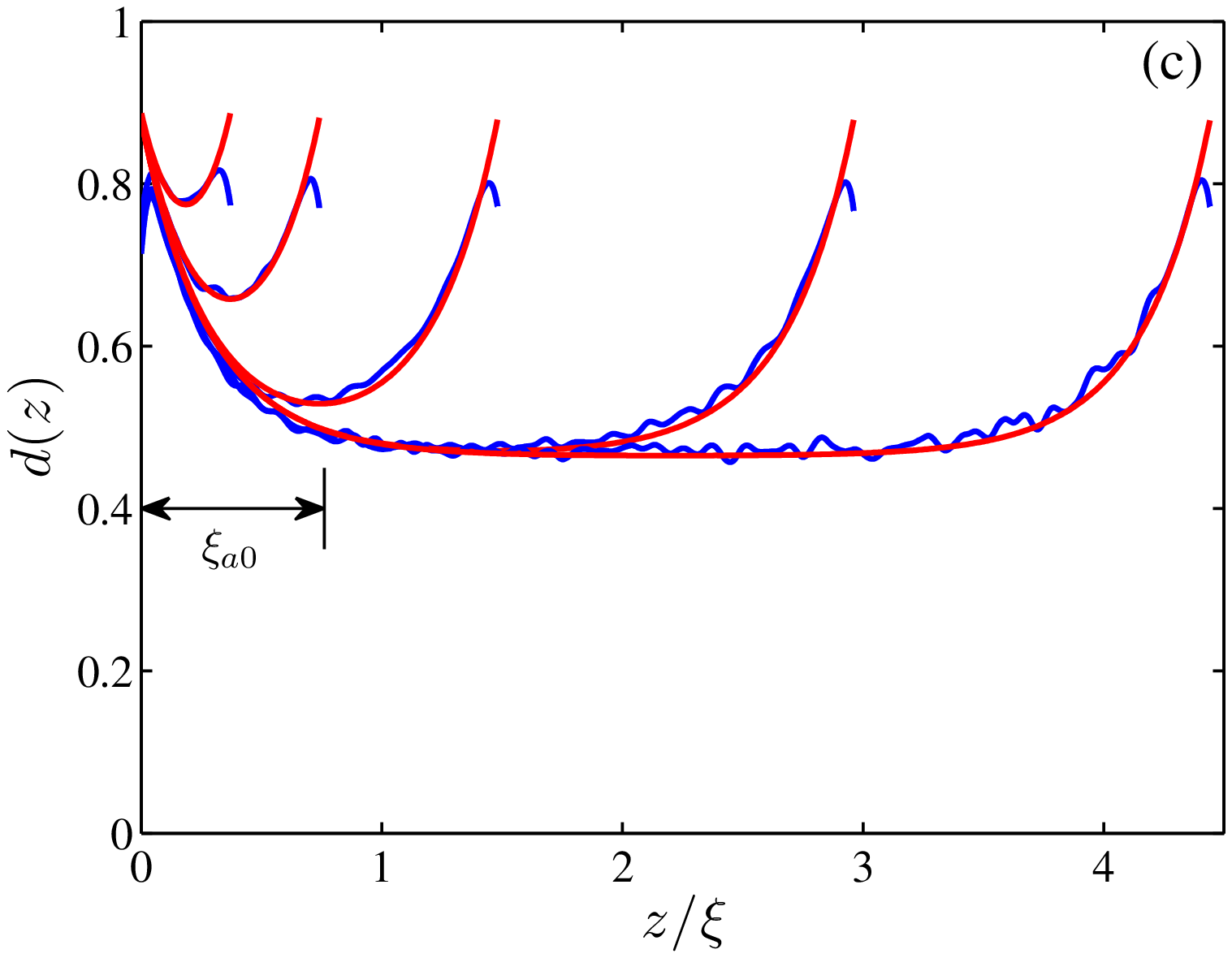}}
\caption{\label{fig:Dofz} 
Comparison between position-dependent diffusion coefficient $d(z)=D(z)/D_0$ found from ab-initio numerical simulations (c.f. Eq.~(\ref{eq:Dofz_definition}), blue curves) and self-consistent theory (Eqs.~(\ref{eq:sceq1}-\ref{eq:bc}), red curves). 
(a) In passive systems ($L/\xi=0.4,\ 0.7,\ 1.5,\ 3.0$ and $4.4$) the diffusion coefficient diminishes in the interior of the system due to the enhanced return probability and, hence, stronger localization correction caused by wave interference. 
(b) For fixed length ($L/\xi=4.4$), an increase of absorption suppresses the localization corrections to the position-dependent diffusion coefficient. The five curves correspond to $L/\xi_{a0}=0.0,\ 3.3,\ 5.7,\ 9.8$ and $21$. 
(c) When absorption ($\xi/\xi_{a0}=1.3$) is added to the five samples shown in (a), the position-dependent diffusion coefficient no longer decreases below its saturated plateau value $D_p$, see Eq.~(\ref{eq:d_p}). In all cases SCT agrees well with the ab-initio simulation of wave transport in disordered waveguides. The cause of the small deviations are discussed in the text.
}
\end{figure}

First, we consider the effect of localization on the local diffusion coefficient as defined by Eq.~(\ref{eq:Dofz_definition}). Fig.~\ref{fig:Dofz}a compares the ab-initio $D(z)$ (blue curves) and the prediction of SCT from Eqs.~(\ref{eq:sceq1}-\ref{eq:bc}) (red curves) in passive systems (no absorption). The behavior of the systems is determined by the ratio of system length $L$ to localization length $\xi$. The five samples shown in Fig.~\ref{fig:Dofz}a correspond to $L/\xi=0.4,\ 0.7,\ 1.5,\ 3.0$ and $4.4$ covering both $L<\xi$ and $L>\xi$ regimes. We observe that without absorption SCT agrees well with the ab-initio simulations. The agreement is achieved for the same values of parameters $\ell$ and $z_0$ found via the procedure described in Sec.~\ref{sec:model}. We would like to comment on two kinds of systematic discrepancies between the numerical data and SCT. (i) We attribute the discrepancy in the boundary regions $0<z\lesssim\ell$ and $L-\ell\lesssim z<L$ to the failure of the diffusion approximation at the boundaries. This is a well known effect which requires a more sophisticated description, such as that by the Milne equation\cite{2007_Akkermans_book}, even without localization corrections. (ii) The second discrepancy arises at $z\sim L/2$ deep in the localization regime $L\gg\xi$. In the most localized system ($L/\xi=4.4$) accessible to us computationally, we find that SCT systematically underestimates the value of the position-dependent diffusion coefficient\cite{2010_Payne_PRL}. An explanation was provided in the work of Tian et al\cite{2010_Tian_PRL} who pointed out the inability of SCT to describe the effects of resonant tunneling which becomes the dominant mechanism of wave transport in a localized system. Based on our simulation, we find significant deviations in systems where $D(z)/D_0$ falls below the value of $\simeq 0.2$, which also agree with the numerical simulations reported in Ref.~\cite{2010_Tian_PRL}.  

Absorption has long been considered a nuisance as it tends to both mask and suppress the effects of wave localization\cite{1984_John_prl,1998_Brouwer,2000_chabanov_nature,2009_Lagendijk_PT}. Indeed, absorption suppresses the contribution of long propagation paths to the return probability which are responsible for localization effects. In Fig.~\ref{fig:Dofz}b we observe an increase of $D(z)/D_0\rightarrow 1$ with a decrease of $\xi_{a0}$ (stronger absorption) while the sample size $L$ is fixed. The suppression occurs first in the middle of the sample (farthermost from the boundaries) where the longest paths are the most probable. We note the development of a plateau $D(z)\simeq D_p$ in the spatial region $\xi_{a0}\lesssim z\lesssim L-\xi_{a0}$. This can be understood based on the following argument. Due to the presence of absorption, a wave that originates at position $z$ in the plateau region can explore only the interval from $z-\xi_{a0}$ to $z+\xi_{a0}$ which does not include a boundary. Hence, as far as the return probability in Eq.~(\ref{eq:sceq2}) is concerned, the effective system size\cite{1998_Brouwer} $L_{eff}\sim 2\times\xi_{a}=2\times\sqrt{D_p\tau_a}$. This argument can be made more quantitative by neglecting the $\partial d(\zeta)/\partial\zeta$ term in Eq.~(\ref{eq:sceq1}). We solve the self-consistent Eqs.~(\ref{eq:sceq1}-\ref{eq:bc}) under the above assumption and obtain the following self-contained equation on $d_p\equiv D_p/D_0$:
\begin{equation}
d_{p}^{-1}= 1+\left(\xi_{a0}/\xi\right)d_{p}^{-1/2}, 
\label{eq:d_p}
\end{equation}
which, in the limit of strong absorptions $\xi_{a0}/\xi\rightarrow 0$, yields an even simpler result $d_{p}\simeq \left[1+\xi_{a0}/\xi\right]^{-1}$. In this expression the parameter $g_{eff}\equiv\xi/\xi_{a0}$ controls the extent of renormalization of the diffusion coefficient $D_p$, similar to the role played by the conductance $g_0\simeq\xi/L$ in passive systems\cite{2010_Payne_PRL}. 

A careful comparison between between the numerical results and SCT in Fig.~\ref{fig:Dofz}b shows a new kind of discrepancy which arises in the localized systems with weak absorption.  We observe that unlike SCT, the ab-initio $D(z)$ becomes asymmetric with respect to the middle of the sample $D(z)\neq D(L-z)$. We hypothesize that this occurs because the absorption tends to affect the energy density inside a localized system differently depending on the position of the localization center in a transmission measurement\cite{2010_Payne_TE}. The detailed study of these weakly absorbing systems goes beyond the scope of this work and will be reported elsewhere.

\subsection{Minimum diffusion coefficient in absorbing systems}

\begin{figure}
\centering{\includegraphics[width=3in,angle=-0]{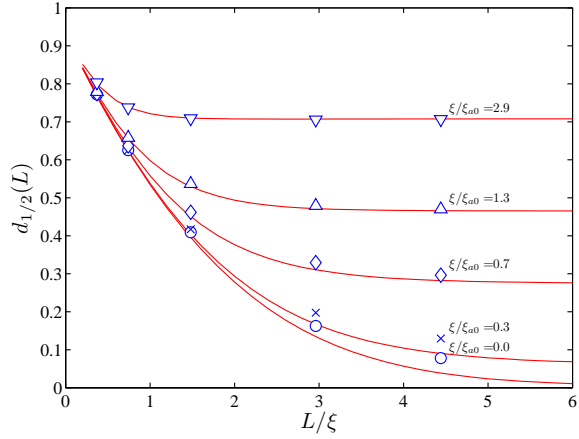}}
\caption{\label{fig:D_middle} 
Existence of the minimum diffusion coefficient is seen from the evolution of  $d_{1/2}\equiv D(z=L/2)/D_0$ with the increase of the system size. In the passive systems ($\xi/\xi_{a0}=0$, open circles) the limit is expected to be zero. In absorbing systems ($\xi/\xi_{a0}=0.3,\ 0.7,\ 1.3,\ 2.9$ shown as cross, diamond, upward and downward triangle symbols respectively) saturation corresponds to formation of the plateau region seen in Figs.~\ref{fig:Dofz}b,c. The saturation value $D_p$ increases monotonically with an increase of $\xi/\xi_{a0}$. Five solid lines are obtained from the self-consistent theory Eqs.~(\ref{eq:sceq1}-\ref{eq:bc}) for each value of the absorption strength.  Qualitative prediction of the minimum value of position-dependent diffusion coefficient in SCT is supported by the numerical simulations. The agreement is  also  quantitative for $D_p/D_0 \gtrsim 0.2$. }
\end{figure}

As seen in Fig.~\ref{fig:Dofz}a, in the passive systems the value of the diffusion coefficient in the middle of the sample $D(z=L/2)$ approaches zero with an increase of $L$. As the boundaries are further removed from $z=L/2$, it becomes progressively more difficult for a wave to escape and the return probability keeps increasing.  As discussed above, the same argument no longer applies in an absorbing random medium. In Fig.~\ref{fig:Dofz}c we plot the position-dependent diffusion coefficient in absorbing ($\xi/\xi_{a0}=1.3$) systems of different size. Unlike passive systems in Fig.~\ref{fig:Dofz}a, the diffusion coefficient stops decreasing after the system size $L$ exceeds the length $\sim 2\xi_{a0}$ and develops a plateau where it is no longer sensitive to the boundaries of the system.

Fig.~\ref{fig:D_middle} shows saturation of the diffusion coefficient in the middle of the sample $D(z=L/2)$ with an increase of the sample size $L$. Importantly, the saturated value $D_p$ is determined by the ratio $\xi/\xi_{a0}$, see Eq.~(\ref{eq:d_p}). As we discussed above, SCT underestimates the value of the diffusion coefficient when the latter becomes small. However, the phenomenon of saturation is observed in both ab-initio simulations and SCT. 

\subsection{Universal scaling of position-dependent diffusion}

\begin{figure}
\vskip 0cm
\centering{\includegraphics[width=3in,angle=-0]{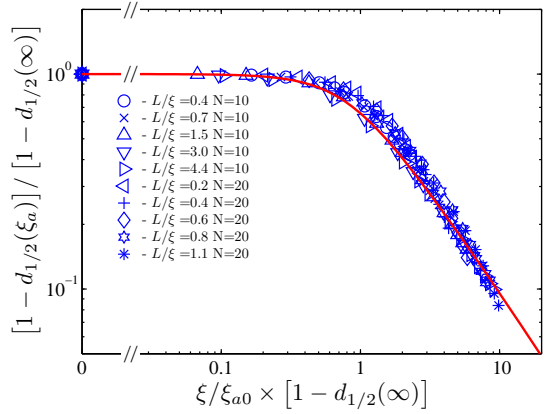}}
\caption{\label{fig:scaling} 
Universal scaling of the local diffusion coefficient in the middle of the sample $d_{1/2} \equiv D(z=L/2)/D_0$ is described by a single parameter. The ab-initio numerical simulations in disordered waveguides with absorption (symbols) agree well with Eq.~(\ref{eq:d_p}), which we derived based on SCT.}
\end{figure}

Single parameter scaling\cite{1979_Anderson} is a key concept in mesoscopic physics. It predicts that when the size of the system increases, the evolution of the average conductance, its variance\cite{1987_Lee_Stone} as well as its entire distribution\cite{1980_Anderson} are determined by the average conductance. In other words, the mesoscopic electronic transport through a disordered system is universal\cite{1991_Altshuler} -- independent of the microscopic parameters of disorder. The concept of universality can be directly applied also to the electromagnetic wave transport\cite{1999_van_Rossum}. Recently, it was shown that in the absence of absorption the concept of position-dependent diffusion coefficient is also universal\cite{2010_Payne_PRL,2010_Tian_PRL,2013_Yamilov_Closed_Channels} -- the shape of $D(z)$ is uniquely determined by the value of conductance.

In absorbing systems, there appears the second parameter related to the absorption length scale\cite{1994_Freilikher_absorption,1998_Brouwer,2001_Deych_sps_abs,2010_Yamilov_Regimes}. Throughout this work we use $\xi/\xi_{a0}$ as such a parameter. The agreement between the numerical simulations and SCT demonstrates that the position-dependent diffusion coefficient is uniquely determined by two dimensionless parameters $g_0\simeq \xi/L$ (related to conductance in passive systems) and $\xi/\xi_{a0}$ -- the only parameters in self-consistent Eqs.~(\ref{eq:sceq1}-\ref{eq:bc}). The parameter $z_0$ entering into Eq.~(\ref{eq:bc}) is not essential as it affects the behavior of $D(z)$ only in the boundary regions where its behavior is not universal\cite{2013_Yamilov_Closed_Channels}.

Eq.~(\ref{eq:d_p}) describing the plateau value of the diffusion coefficient in absorbing systems depends only on one parameter $\xi/\xi_{a0}$. This suggests a new universal behavior. In Fig.~\ref{fig:scaling} we compare this prediction of SCT (solid line) to our numerical results in different disordered absorbing waveguides (symbols) and observe a good agreement.

\section{Conclusions\label{sec:conclusions}}

In this work we compared the results of the ab-initio numerical simulation of wave transport through disordered absorbing waveguides to the prediction of self-consistent theory with position-dependent diffusion. This theory has the power to provide a simple intuitive description of an intricate interplay between localization effects due to wave interference and absorption. The following results were obtained. (i) We demonstrated that in localized regime ($L>\xi$) for weak/strong absorption SCT provides the correct qualitative/quantitative description of the renormalizations in position-dependent diffusion coefficient. (ii) We pointed out the key distinction in scaling of the position-dependent diffusion in passive and absorbing media. In the former, the diffusion coefficient vanishes in the limit of infinite size. In contrast, in absorbing random media, the local diffusion coefficient always remains finite so that true localization in the sense of vanishing diffusion is never reached. This conclusion appear to be borne out also by supersymmetric field theory\cite{2013_Tian}.

We would like to point out that a similar (but not equivalent) effect occurs in the mesoscopic electronic systems due to the effects of dephasing\cite{2007_Akkermans_book} which introduces the cutoff length scale beyond which waves no longer contribute coherently to the return probability. Instead, these inelastically scattered (de Broglie) waves of the electrons contribute to the incoherent background which only obscures the localization phenomena. In contrast, optical absorption removes photons while the remaining component retains all information about interferences. This observation points to the possibility of observation of the phenomenon of position-dependent diffusion in optical system that we discuss next.

There are three main conditions that would allow the direct experimental observation position-dependent diffusion. First, without absorption the system must be in the localization regime in the limit of infinitely large size. Second, the absorption, or more generally loss, cannot be too strong so that the localization effects are not completely washed out. Third, because the renormalization of diffusion occurs away from the boundaries, the interior of the medium has to be experimentally (non-invasively) monitored. These conditions can be met in the two dimensional disordered optical waveguides where the position-dependent diffusion is indeed observed\cite{2013_Yamilov_Dofz_experiment}. 


\section{Acknowledgements}

We gratefully acknowledge the discussions with H.~Cao, R.~Sarma, A.~D.~Stone and B.~Shapiro. We also acknowledge comments by S.~Skipetrov, P.~Sebbah during the early stages of this work. We are thankful to C.~Tian for reading and commenting on the manuscript. Financial support was provided by National Science Foundation under grants Nos. DMR-0704981 and DMR-1205223. Computational resources were provided under the Extreme Science and Engineering Discovery Environment (XSEDE) grant No. DMR-100030.

\end{document}